\documentclass[conference]{IEEEtran}
\IEEEoverridecommandlockouts
\usepackage{cite}
\usepackage{pgfplots}
\usepackage{pgfplotstable} 
\pgfplotsset{compat=1.18}
\usepackage{amsmath,amssymb,amsfonts}
\usepackage{soul}
\usepackage{algorithmic}
\usepackage{graphicx}
\usepackage{textcomp}
\usepackage{xcolor}
\usepackage{amsfonts}
\usepackage{booktabs}
\usepackage{float}
\usepackage{hyperref}
\usepackage{graphicx}
\usepackage{grffile}
\usepackage{longtable}
\usepackage{tabularx}
\usepackage{makecell}
\usepackage{wrapfig}
\usepackage{gensymb}
\usepackage{textcomp}
\usepackage{multicol}
\usepackage{multirow}
\usepackage[table]{xcolor}
\usepackage{xcolor}
\usepackage{amsmath}
\usepackage{algorithm}
\usepackage{verbatim}
\usepackage{graphicx}
\usepackage{caption}
\usepackage{listings}
\usepackage{array}
\usepackage{subcaption}
\usepackage{booktabs}
\usepackage{amsmath}
\usepackage{enumitem}
\usepackage{dblfloatfix}
\usepackage{balance}
\usepackage{lastpage}
\usepackage{lipsum}
\def\arraystretch{1.25}
\usepackage[utf8]{inputenc}

\begin{document}
\title{Microbenchmarking Cloud Cryptographic Workloads for Privacy-Preserving Healthcare IoT}

\author{\IEEEauthorblockN{Jeremiah L. Webb\IEEEauthorrefmark{1}, Laxima Niure Kandel\IEEEauthorrefmark{2}, Deepti Gupta\IEEEauthorrefmark{3}, Lavanya Elluri\IEEEauthorrefmark{4}}
\IEEEauthorblockA{\IEEEauthorrefmark{1}\IEEEauthorrefmark{2}{Dept. of Electrical Engineering and Computer Science, Embry–Riddle Aeronautical University, Daytona Beach, Florida, USA}\\\IEEEauthorrefmark{3}\IEEEauthorrefmark{4}{Dept. of Computer Information Systems, Texas A\&M University - Central Texas, Texas, USA}\\}
\IEEEauthorrefmark{1}webbj31@my.erau.edu, 
\IEEEauthorrefmark{2}niurekal@erau.edu,
\IEEEauthorrefmark{3}d.gupta@tamuct.edu,
\IEEEauthorrefmark{4}elluri@tamuct.edu}
\maketitle

\begin{abstract}
Cryptographic operations are an essential component of cloud security architectures; their comprehensive performance characterization across different cloud services, hardware architectures, and programming language implementations remains unknown. Specifically, healthcare IoT devices are highly vulnerable and frequently targeted, yet the cryptographic performance trade-offs in their cloud security architectures remain poorly understood. This research presents an extensive microbenchmark study evaluating the performance of core cryptographic workloads—including SHA HMAC generation, AES encryption/decryption, Elliptic Curve Cryptography (ECC) signature generation and verification, and RSA encryption/decryption, across Function-as-a-Service (FaaS) integrated with Key Management Services (KMS) from Amazon Web Services (AWS) and Microsoft Azure. We evaluate FaaS platforms using Elastic Compute Cloud (EC2) instances and Azure Virtual Machines, specifically using burst-optimized instance types to analyze performance under typical cloud workload patterns. The benchmark encompasses a comprehensive multi-dimensional analysis spanning two CPU architectures (x86\_64 and Arm64), six widely adopted programming languages (Rust, Go, Python, Java, C\#, and TypeScript), multiple memory allocation configurations, and diverse instance types to capture the complex interplay between these factors. This study identifies optimal configurations for cryptographic workloads in FaaS environments, improving performance and cost efficiency while enabling secure and timely data protection for healthcare IoT applications.

\end{abstract}

\begin{IEEEkeywords}
Amazon Web Services, Azure, cloud computing, cryptography, function-as-a-service, microbenchmark, performance analysis.
\end{IEEEkeywords}

\section{Introduction}
\label{sec:introduction}

Cloud computing has become a key enabler for modern applications by providing scalable, on-demand access to computing resources with a pay-as-you-go model \cite{b1}\cite{b3}. In particular, healthcare IoT applications, such as Remote Patient Monitoring (RPM), continuously generate sensitive physiological data that must be securely transmitted and processed in the cloud to enable real-time analytics such as anomaly detection, and predictive decision-making. To support scalability, rapid deployment, and reduced operational overhead, these applications increasingly rely on Function-as-a-Service (FaaS) platforms (e.g., AWS Lambda and Azure Functions), which offer event-driven execution and automatic scaling without requiring infrastructure management.

Ensuring the security and privacy of healthcare data is critical due to strict regulatory requirements such as HIPAA. Cloud-based healthcare IoT systems depend heavily on cryptographic operations, including AES, RSA, ECC, SHA, and HMAC, to guarantee data confidentiality, integrity, and authenticity. These operations are commonly integrated with cloud-native Key Management Services (KMS) for secure key generation, storage, and usage. However, cryptographic operations are computationally intensive, and in FaaS environments, their performance directly affects execution latency, cost (due to per-invocation billing), and system responsiveness.

Despite their importance, there is limited understanding of how cryptographic workloads behave in serverless (FaaS) environments across different cloud providers, programming languages, CPU architectures (e.g., x86\_64 vs. Arm64), and memory configurations. The dynamic nature of FaaS, characterized by cold starts, resource constraints, and ephemeral execution, introduces additional variability that can significantly impact cryptographic performance. Existing studies often rely on synthetic workloads or do not evaluate real cryptographic operations with integrated KMS, making their findings less applicable to real-world, security-critical applications like healthcare IoT. This gap is particularly critical for healthcare scenarios, where delays in encryption, decryption, or signature verification can impact timely detection of critical events and overall system reliability. Moreover, the lack of systematic benchmarking makes it difficult for developers to choose optimal configurations that balance security, latency, and cost in serverless environments.

To address this challenge, this research presents a comprehensive microbenchmarking study of core cryptographic operations, including SHA HMAC generation, AES encryption/decryption, ECC signature generation and verification, and RSA encryption/decryption, across FaaS platforms only, specifically AWS Lambda and Azure Functions. Our evaluation spans multiple programming languages (Rust, Go, Python, Java, C\#, and TypeScript), memory configurations, and CPU architectures, with integrated use of cloud-native KMS to reflect realistic deployment scenarios. By systematically analyzing execution latency, throughput, and resource behavior, this work provides practical, evidence-based insights for optimizing cryptographic workloads in serverless environments.

To guide this study, we investigate the following research questions:
\begin{itemize}
\item \textbf{RQ1:} How do cryptographic workloads perform across AWS and Azure FaaS platforms under different configurations (programming language, CPU architecture, and memory allocation)?
\item \textbf{RQ2:} What configuration choices most significantly impact the latency and efficiency of cryptographic operations in serverless environments with integrated KMS?
\end{itemize}

Key contributions of this research are as follows:
\begin{itemize}
\item We conduct a comprehensive analysis of the implementation cryptographic workloads across multiple programming languages and configurations in AWS and Azure, using FaaS paradigms.
\item We propose a custom methodology for microbenchmarking cryptographic workloads within these cloud environments.
\item We present detailed microbenchmark results to evaluate the throughput efficiency and execution speed across various configurations to identify optimal deployment strategies for the cryptographic workloads.
\end{itemize}

The remainder of this paper is organized as follows: Section~\ref{sec:relatedwork} reviews related work and establishes the current state of research on microbenchmarking cloud FaaS platform and healthcare domain. Section~\ref{usecase} presents threats on RPM ecosystem. Section~\ref{experimentsetup} presents a comprehensive overview of our experimental design and the implementation of cryptographic workloads across cloud environments. Section~\ref{benchmark} describes our custom microbenchmarking framework and architecture for evaluating cryptographic operations on AWS and Azure FaaS platform. Section~\ref{results} presents our experimental results, analyzing performance patterns and identifying optimal configurations for execution speed and throughput efficiency across different use cases. Finally, Section~\ref{Conclusion} provides concluding remarks, discusses limitations, and outlines directions for future work.

\section{Background and Related Work}

\label{sec:relatedwork}
Microbenchmarking tests specific small portions of an application's performance. Researchers have developed various benchmarking frameworks and methods to evaluate performance in multiple cloud environments and configurations \cite{b8}\cite{b18}\cite{b19}\cite{b15}. Numerous studies have examined the performance and cost differences among various cloud providers. For example, Chaisiri et al. (2011) introduced a formal mathematical decision model to assist in selecting cloud computing services within a multisourcing environment, considering both cost and risk factors \cite{b8}. Srisakthi et al. (2018) developed a Secure Encryption Model (SEM) using Hadamard Transforms for secure cloud data storage. Their benchmarks demonstrated that compared to elliptic curve cryptography (ECC), encryption using Hadamard transforms reduced storage requirements by 10\% and decreased encryption execution time by 25\% \cite{b16}. AWS also provides an open-source tool, AWS Lambda Power Tuning, that helps developers optimize memory sizes for AWS Lambda functions, this tool automates evaluations by testing various memory sizes and measuring performance metrics such as execution time and cost \cite{b18}.
\subsection{Key Management Service}
NIST defines key management as managing cryptographic keys throughout their life cycle, including secure generation, storage, distribution, usage, and destruction \cite{b4}. An automated system that performs these functions is known as a cryptographic key management system \cite{b4}. A comparative study of Key Management Services (KMS) from AWS, Google Cloud, and Oracle Cloud using load, stress, and benchmark testing showed that AWS outperformed competitors \cite{b9}. AWS achieved lower error rates, reduced response times, lower latency, and consistently higher throughput for AES256 encryption and decryption \cite{b9}. The results underscore the importance of security practices, such as data encryption, in addressing challenges like key loss in cloud environments. However, this testing only utilized JMeter with HTTPS endpoints of the key management services \cite{b9}, not language-specific libraries on FaaS.

\subsection{Function-as-a-Service}
FaaS is a serverless computing model that enables developers to create applications with event-driven, stateless code executed in the cloud \cite{b2}. By eliminating the need to provision cloud instances, serverless computing ensures high availability and scalability, allowing developers to concentrate on the core application logic using established cloud services \cite{b11}. Azure’s consumption plan does not permit the configuration of different memory sizes and computing architectures during deployment \cite{b12}\cite{b13}. Researchers have developed several frameworks to benchmark FaaS; one notable example is FunctionBench, which currently benchmarks AWS, Google, and Azure for AES256 without using key management services \cite{b14}. Copik et al. (2021) introduced the Serverless Benchmark Suite (SeBS), which provides a standardized methodology to compare serverless providers. SeBS evaluates FaaS performance, reliability, and suitability across various workloads, including compression and multimedia processing, and analyzes the impacts of cold and warm invocation across programming languages \cite{b19}. Additionally, AWS and Cascadeo (2023) conducted experiments with nearly 2,000 Lambda functions across three runtimes (Python, NodeJS, Ruby), various configurations, memory sizes, worker threads, and workloads such as compression, SHA256 hashing, and in-memory sorting. The findings revealed significant performance and cost improvements by migrating to Arm-based Lambdas, especially for CPU-intensive workloads \cite{b23}.

\subsection{Benchmark Testing}
Benchmark testing is a performance testing method that measures and compares system performance through standardized tests \cite{b33}. Its primary objective is to provide an objective and consistent approach for assessing and comparing computer systems’ performance and identifying bottlenecks for improved efficiency \cite{b72}. This approach is widely employed across fields such as computer architecture, hardware and software design, and system optimization \cite{b33}. Microbenchmarks typically run repeatedly for a set duration (e.g., 1 second) to measure the average execution time, reporting the results based on multiple iterations (e.g., 20) \cite{b48}. Software microbenchmarks can be viewed as performance unit tests for microservices \cite{b49}. Particularly in cloud environments, microbenchmarking with infrastructure-as-code configurations is considered a best practice for reproducibility \cite{b36}\cite{b35}.

\subsection{Cryptographic Workloads for Healthcare IoT Applications}

This research~\cite{reis2021cryptography} investigated Web Assembly and JavaScript solutions that enable client-side cryptography in web applications and compares their performance against server-side cryptography. The authors contextualized the study within two healthcare web applications: a prototype for patient record sharing during acute stroke care and an application for sharing data in sleep medicine treatment.  The authors~\cite{aledhari2017new} discussed the design and implementation of a hybrid real-time cryptography algorithm to secure lightweight wearable medical devices. The proposed system is based on an emerging innovative technology between the genomic encryptions and the deterministic chaos method to provide a quick and secure cryptography algorithm for real-time health monitoring that permits for threats to patient confidentiality to be addressed. The authors~\cite{lakhan2021hybrid} proposed a new healthcare architecture for workflow applications based on heterogeneous computing nodes layers: an application layer, management layer, and resource layer. Golec et al. (2023) introduced HealthFaaS \cite{healthFaaS} , a novel framework that shifts cardiovascular monitoring from traditional cloud servers to a serverless computing model. By utilizing Google Cloud Functions, the authors demonstrated that a FaaS approach significantly reduces operational costs and infrastructure management while maintaining high scalability for real-time patient data processing. Several prior works have contributed to advancements in intelligent and secure computing systems~\cite{karimagent, kavuriagentic, kavuri2025securefed, nair2025androids, gupta2021hierarchical}.

\subsection{Research Gap}

Despite advances in cloud performance benchmarking, significant gaps remain in understanding cryptographic workload behavior in healthcare IoT environments. Existing studies either exclude cloud-native KMS integration \cite{b14}, evaluate KMS only through HTTP endpoints rather than language-specific implementations \cite{b9}, or focus on general workloads that do not reflect the latency-sensitive and security-critical nature of healthcare IoT data \cite{b19}\cite{b23}. No prior work systematically compares cryptographic performance across cloud providers, FaaS models, CPU architectures, programming languages, and configurations with integrated KMS in the context of real-time medical data processing and compliance requirements (e.g., HIPAA). This gap limits the ability of healthcare systems to make evidence-based deployment decisions that balance security, latency, and cost for patient-centric applications. Additionally, the lack of reproducible benchmarking frameworks tailored to healthcare IoT restricts validation, trust, and adaptability as cloud and edge platforms evolve.

\begin{figure*}[ht!]
    \centering
    \includegraphics{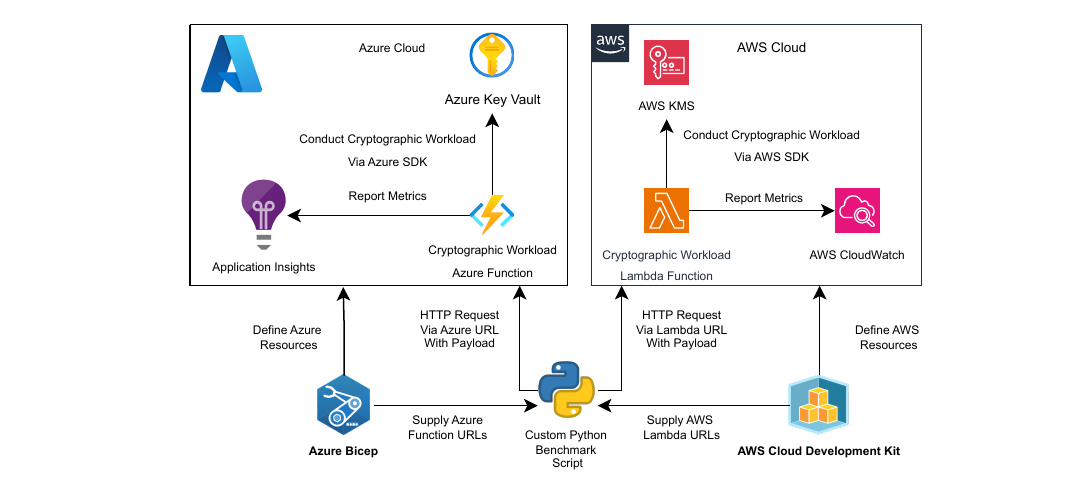}
    \caption{Azure and AWS FaaS microbenchmark architecture.} 
    \label{fig1}
\end{figure*}

\section{Use Case: Secure Remote Patient Monitoring in Healthcare IoT}
\label{usecase}
Remote Patient Monitoring (RPM) is an important system of modern healthcare IoT, enabling continuous monitoring of patients outside traditional clinical settings. In this use case, wearable and home-based IoT devices, such as smartwatches, ECG monitors, pulse oximeters, and blood pressure sensors, collect real-time physiological data from patients with chronic conditions (e.g., cardiovascular diseases, diabetes, or respiratory disorders). These devices transmit sensitive health data to cloud-based platforms for storage, processing, and analysis, where clinicians and intelligent systems can detect anomalies and trigger timely interventions. Given the highly sensitive nature of medical data, RPM systems rely extensively on cryptographic operations to ensure data confidentiality, integrity, and authenticity. Data generated at the device level is encrypted using symmetric algorithms such as AES before transmission. Secure communication channels (e.g., TLS) leverage asymmetric cryptography (e.g., RSA or ECC) for key exchange, while HMAC-based mechanisms ensure message integrity. Additionally, cloud-native Key Management Services (KMS) are used to securely generate, store, and manage cryptographic keys, ensuring compliance with healthcare regulations such as HIPAA. In this architecture, serverless FaaS platforms (e.g., AWS Lambda, Azure Functions) are commonly used to process incoming data streams due to their scalability and event-driven nature.

However, this architecture introduces several security and performance challenges. First, the continuous and latency-sensitive nature of RPM requires cryptographic operations within FaaS environments to be highly efficient. Any delay in encryption, decryption, or signature verification caused by cold starts, limited memory allocation, or inefficient runtime configurations can delay real-time analytics, potentially impacting timely detection of critical health events such as arrhythmias or oxygen desaturation. Second, IoT devices are resource-constrained, often limiting their ability to perform computationally expensive cryptographic operations, which may lead to weaker implementations or increased reliance on cloud-side processing.

From a threat perspective, RPM systems are vulnerable to data interception attacks, where unencrypted or weakly encrypted data can be exposed during transmission. Man-in-the-middle (MITM) attacks can compromise communication channels if key exchange mechanisms are not properly implemented. Additionally, improper integration with cloud-based KMS may introduce key management vulnerabilities, such as unauthorized access to cryptographic keys or increased latency due to frequent key retrieval operations within serverless workflows. Variability in FaaS configurations—across cloud providers, CPU architectures (x86\_64 vs. Arm64), programming languages, and memory settings can further lead to inconsistent cryptographic performance and potential misconfigurations, increasing the attack surface. Another critical challenge is the trade-off between security and performance in serverless environments. For example, certain cryptographic operations (e.g., ECC vs. RSA) may perform differently depending on the runtime environment, directly affecting system responsiveness. Without a systematic understanding of these trade-offs, healthcare organizations may deploy suboptimal configurations that either compromise security or fail to meet real-time performance requirements.

The proposed work directly addresses these challenges by providing a comprehensive benchmarking framework for evaluating cryptographic workloads across FaaS platforms, architectures, and programming environments with integrated KMS. By offering empirical insights into latency, throughput, and resource utilization, this approach enables healthcare system designers to make evidence-based decisions when selecting cryptographic algorithms, runtime environments, and configurations. For instance, it can guide which programming language or memory configuration minimizes latency for encryption tasks, or how KMS integration impacts response time in serverless pipelines.

Ultimately, this work supports the development of secure, efficient, and scalable RPM systems, ensuring that patient data remains protected while enabling real-time clinical decision-making. By reducing performance bottlenecks and minimizing security risks, it enhances the reliability of healthcare IoT applications and contributes to improved patient outcomes in distributed care environments.

\section{Experiment Setup}
\label{experimentsetup}
The experimental setup for this research involved two cloud providers: AWS and Azure. Each provider underwent microbenchmark testing to measure and compare system performance in executing cryptographic workloads using standardized tests. We selected the most widely used programming languages for this research, as listed in Tables \ref{tab:faas_grouped} and \ref{tab:iaas_grouped}. Both AWS and Azure provide official SDKs for multiple programming languages \cite{b44}\cite{b45}. 

\begin{figure}[ht!]
    \centering
    \includegraphics[width=\columnwidth]{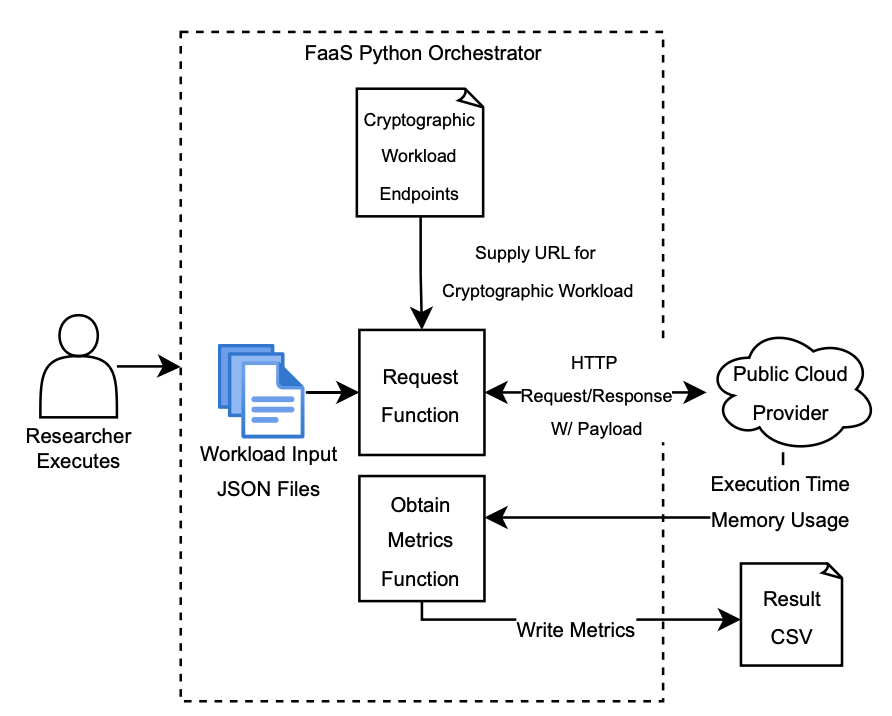}
    \caption{The flow of execution and data collection for FaaS microbenchmarks}
    \label{fig2}
\end{figure}

Additionally, all microbenchmark tests conducted during this research used a pre-generated 4-kilobyte JSON-formatted string payload for cryptographic workloads across FaaS tests. For the experimental setups, refer to Figure \ref{fig1}, Figure \ref{fig2} for the orchestration of the FaaS microbenchmarks. The source code for this experiment is available for public review in a GitHub repository. For the purpose of blind review, the link is currently omitted and will be included in the final camera-ready version\footnote{\url{https://github.com}}.

\begin{table}[h]
    \caption{Tested Cryptographic Algorithms and Workloads Supported by AWS and Azure by their respective KMS service}
    \centering
    \begin{tabular}{|p{60pt}|p{95pt}|c|c|}
        \hline
        \centering{\textbf{Algorithm}} & \centering{\textbf{Cryptographic Workload}} & \textbf{Azure} & \textbf{AWS} \\
        \hline
        \centering{HMAC-SHA-256} & HMAC Generation &  & X \\
        \hline
        \centering{HMAC-SHA-384} & HMAC Generation &  & X \\
        \hline
        \centering{AES-256 GCM} & Symmetric \newline Decryption/Encryption &  & X \\
        \hline
        \centering{ECC (P-256)} & Signature \newline Generation/Verification & X &  X \\
        \hline
        \centering{ECC (P-384)} & Signature \newline Generation/Verification & X & X \\
        \hline
        \centering{AES-256 CTR} & Symmetric \newline Encryption/Decryption \newline Used in Envelope Encryption (OpenSSL Libraries) & X & X \\
        \hline
        \centering{RSA 2048-bit} \newline RSAES-OAEP-SHA-256 & Envelope \newline Encryption/Decryption & X & X \\
        \hline
        \centering{RSA 3072-bit} \newline RSAES-OAEP-SHA-256 & Envelope \newline Encryption/Decryption & X & X \\
        \hline
        \centering{RSA 4096-bit} \newline RSAES-OAEP-SHA-256 & Envelope \newline Encryption/Decryption & X & X \\
        \hline
    \end{tabular}

    \label{tab:cryptoalgoworkloads}
\end{table}

\subsection{Cryptographic Workloads}
The selected workloads reflect common cryptographic operations used in cloud-based key management and security services. These operations secure data both in transit and at rest using AES and RSA \cite{b24}, ensure data integrity with HMAC-SHA (Hash-Based Message Authentication Code) \cite{b24}, and support authentication in cloud environments \cite{b25}. Each algorithm was chosen based its on real-world applicability, computational complexity, and relevance to cryptographic tasks in AWS and Azure environments.

AWS supports various cryptographic algorithms through AWS KMS \cite{b26}. This includes symmetric encryption keys, such as AES-GCM with an HMAC-based extract-and-expand key derivation function \cite{b29}, and asymmetric keys for RSA and ECC. HMAC generation, for data integrity and authenticity, uses cryptographic hash functions for secure message verification \cite{b32}. These options enable encryption, decryption, signing, and verification operations tailored to diverse security requirements. A comprehensive overview of the tested workloads and supported cryptographic algorithms for both the cloud providers is presented in Table \ref{tab:cryptoalgoworkloads}. Developers and other cloud services can access AWS KMS keys via role based access control (RBAC) IAM policies and roles \cite{b84}\cite{b47}.

Azure emphasizes software-protected keys through its Key Vault service \cite{b27}\cite{b31}. Azure Key Vault safeguards cryptographic keys and secrets, allowing secure storage and management of sensitive information such as passwords, API keys, certificates, and cryptographic keys. Integration with Azure services and RBAC via Azure Active Directory (AAD) ensures secure access \cite{b76}. Currently, Azure Key Vault supports RSA and ECC workloads \cite{b27}\cite{b77} only. RSA envelope encryption with AES256-CTR simulates RSA workloads, aligning with AWS and Azure's envelope encryption recommendations \cite{b26}\cite{b30}. Envelope encryption, which involves encrypting data keys using another encryption key (key-encryption or wrapping key), enhances data security by protecting the encryption keys themselves \cite{b26}\cite{b28}\cite{b30}. Refer to Table \ref{tab:cryptoalgoworkloads} for detailed workload information.

\begin{table}[h]
\centering
\caption{Tested AWS Lambda and Azure Functions}
\setlength{\tabcolsep}{3pt}
\renewcommand{\arraystretch}{1}

\begin{tabularx}{\columnwidth}{|c|X|c|c|X|}
\hline
\textbf{Cloud} & \textbf{Languages} & \textbf{Arch} & \textbf{Memory (MB)} & \textbf{Cost*} \\
\hline

\textbf{AWS} &
Java 21, Python 3.11, TypeScript 5.7.2, C\#, Go 1.23, Rust 1.83 &
x86\_64, Arm64 &
128–3008 &
2.1e-9–4.9e-8 (x86) \newline 1.7e-9–3.9e-8 (Arm) \\
\hline

\textbf{Azure} &
Java 21, Python 3.11, TypeScript 5.7.2, C\# &
x86\_64 &
Dynamic (Max 1500) &
1.6e-8 \\
\hline

\end{tabularx}

\vspace{2mm}
{\raggedright \footnotesize *Cost in USD MB/ms as of 2025-03-06 (US-East regions).\par}

\label{tab:faas_grouped}
\end{table}

\subsection{FaaS Experiment}
The FaaS experiment setup utilized infrastructure-as-code (IaC) due to the numerous configurations involved. IaC automates the definition and management of deployment environments, system configurations, and infrastructure through source code \cite{b34}\cite{b36}. Developers define provisioning and deployment via code, managed through software engineering practices like version control, code review, and continuous integration, facilitating rapid and reproducible deployments \cite{b35}\cite{b36}. Both Azure and AWS FaaS platforms access cryptographic keys via their respective RBAC policies, Key Vault, and AWS KMS, as recommended by both providers \cite{b46}\cite{b47}. Memory allocation pricing varies by provider and configuration (Table \ref{tab:faas_grouped}) \cite{b70}\cite{b71}.

To deploy AWS Lambda function configurations \footnote{\url{https://github.com}}, the AWS Cloud Development Kit (CDK) was employed due to its multiple memory settings \cite{b60}. AWS CDK, an open-source framework, allows the definition of cloud resources in languages such as TypeScript, JavaScript, Python, Java, and C\# \cite{b37}. CDK promotes infrastructure reuse and consistency across deployments \cite{b36}\cite{b37}. Deployed Lambda functions, packaged via zip files \cite{b58}, receive HTTPS endpoints (Lambda URLs) that trigger execution upon HTTP requests \cite{b38}. All functions were deployed in the US-east-1 region, using environment variables to specify AWS KMS key IDs and ARNs \cite{b63}.

Similarly, Azure Functions configurations\footnote{\url{https://github.com}}  were deployed using Bicep, a simplified DSL for Azure resource deployment \cite{b39}. Azure Functions, also deployed via zip files \cite{b59}, provide HTTP-accessible function URLs. Currently, Azure Functions do not support the Arm64 architecture and impose a 1.5 GB memory limit on the consumption plan \cite{b61}\cite{b62}. Environment variables specify Key Vault key names and IDs \cite{b64}, and Azure Application Insights with Azure Monitor records performance metrics \cite{b78}. Azure Functions were deployed in the East US region.

As illustrated in Figure \ref{fig1}, after obtaining the AWS Lambda and Azure Function URLs through the IaC frameworks, a custom Python script orchestrates microbenchmark testing (Figure \ref{fig2}a). The script sends HTTP requests with payloads to both cloud functions, thereby triggering cryptographic operations. Azure Functions utilize the Azure SDK for cryptographic tasks via Key Vault, reporting metrics to Application Insights. The AWS Lambda functions use the AWS SDK for AWS KMS operations, reporting performance metrics to the AWS CloudWatch. This standardized architecture ensures consistent performance measurements and comparisons across both cloud platforms.

\section{Benchmark Implementation}
\label{benchmark}
This section describes the development of a custom benchmark implementation for both IaaS and FaaS platforms on AWS and Azure. All benchmark implementation code is publicly available on GitHub.\footnote{\url{https://github.com}}\footnote{\url{https://github.com}}.

\subsection{Design - FaaS}
To systematically compare AWS and Azure, a microbenchmark implementation was developed. Figure \ref{fig2}a illustrates the full architecture and orchestration. Python 3.11 was chosen to develop the custom benchmark script due to its simplicity and powerful libraries, particularly the \texttt{requests} library for managing HTTP calls. A randomly generated 4-kilobyte JSON-formatted string serves as the payload across all cryptographic workloads for both cloud providers. Native Python libraries were used to handle JSON file operations.
\begin{table}[h]
   \caption{Local Computer Specification for FaaS orchestration benchmark}
\begin{tabular}{ll}
\hline
\textbf{Operating System}         & Arch Linux x86\_64             \\ \hline
\textbf{Host}                     & 82K8 Legion S7 15ACH6          \\ \hline
\textbf{Kernel}                   & 6.13.2-arch1-1                 \\ \hline
\textbf{Shell}                    & bash 5.2.37                    \\ \hline
\textbf{CPU}                      & AMD Ryzen 7 5800H @ 4.463GHz   \\ \hline
\textbf{GPU}                      & NVIDIA GeForce RTX 3060 Mobile \\ \hline
\textbf{Memory}                   & 15348 (MiB)                    \\ \hline
\textbf{Wired Network Connection} & 500 Mbps Up and Down           \\ \hline
\end{tabular}
\label{localcomptable}
\end{table}
\subsection{Utilization - FaaS}
The orchestration of FaaS microbenchmarks involves running a custom Python script from a local computer, whose specifications are detailed in Table \ref{localcomptable}. Executing the Python script triggers the FaaS configurations for both AWS and Azure, running each workload 30 times to capture both cold and warm starts. Cold starts, occurring on the initial invocation, introduce a performance penalty as the cloud provider initializes resources. Subsequent invocations (warm starts) reuse existing resources to optimize execution performance \cite{b40}. Although cold starts constitute only about 1\% of Lambda executions \cite{b39}, warm starts significantly reduce latency. Each HTTP request includes the cryptographic workload details in the request body, and upon successful execution, the Lambda function returns a response containing the workload results.

\subsection{Recorded Metrics - FaaS}
Metrics collection begins with cold starts on the first iteration, while warm-start metrics are recorded after ten initial executions. AWS Lambda functions report performance data, including execution time in milliseconds and memory usage in MB, directly to AWS CloudWatch, a comprehensive monitoring and observability service for AWS, on-premises, and cloud-based applications \cite{b50}\cite{b51}. The collected data adheres to AWS standard metrics \cite{b52}\cite{b53}. For Azure, we employ Azure App Insights and Azure Monitor for standard request metrics \cite{b78}\cite{b85}.

\section{Results and Findings}
\label{results}
This section presents the results of FaaS microbenchmark tests, all of which achieved a 100\% success rate.

To address RQ1, we compared the performance of cryptographic workloads between AWS and Azure across both FaaS platform. Our analysis identified several key factors influencing execution times, including memory allocation, instance type, architecture (x86\_64 vs. ARM64), and cold-start behavior in FaaS environments. AWS generally exhibited lower execution times in FaaS due to more effective warm-start retention strategies, whereas Azure's performance varied significantly based on programming language configuration. 

To address RQ2, we evaluated how infrastructure-specific optimizations, such as memory configurations, vCPU allocations, and architectural choices, impact performance. Within FaaS environments, increased memory allocation typically improved execution times substantially, though gains diminished beyond a specific threshold. 

Additionally, we assessed cost efficiency across various configurations. Our findings revealed that the most cost-effective FaaS setups involve balancing execution time and memory allocation rather than selecting the highest available memory. These insights offer actionable guidance for developers seeking to optimize cryptographic workloads, balancing both performance and cost.

For cost-efficiency calculations, we assumed that Azure provisions 1024 MB of memory for all workloads. Efficiency was defined as achieving the lowest cost per unit of throughput. The following equations were used to identify the most cost-efficient configuration for each workload:

\begin{equation}
    \text{MeanCost} = \text{MeanExecutionTime}\times\text{CostPerMillisecond}
\end{equation}

\begin{equation}
    \text{RequestsPerSecond} = \frac{1}{\text{MeanExecutionTime}}
\end{equation}

\begin{equation}
    \text{CostEfficiency} = \frac{\text{RequestsPerSecond}}{\text{MeanCost}}
\end{equation}

Overall, the results confirm that previous benchmark studies are consistent: across all workloads, higher memory allocation decreases the execution time of the workload, regardless of language \cite{b79}\cite{b28}, as anticipated from theory. To facilitate figure interpretation, Table \ref{tab:colorlegend} provides a legend for the stacked bar colors in performance charts.

\begin{table}[h]
    \caption{Cold and Warm Start Color Legend (Languages In Order)}
    \centering
    \renewcommand{\arraystretch}{1.5}
    \begin{tabular}{|c|c|c|c|c|c|c|c|}
        \hline
        & C\# & Go & Java & Python & Rust & TypeScript \\ 
        \hline
        \textbf{Warm} & \cellcolor[HTML]{5A2073} & \cellcolor[HTML]{00ADD8} & \cellcolor[HTML]{F80000} & \cellcolor[HTML]{FFD43B} & \cellcolor[HTML]{D66B00} & \cellcolor[HTML]{3178C6} \\ 
        \hline
        \textbf{Cold} & \cellcolor[HTML]{9931c5} & \cellcolor[HTML]{68d9fc} & \cellcolor[HTML]{f76d6d} & \cellcolor[HTML]{C59C4F} & \cellcolor[HTML]{E08A39} & \cellcolor[HTML]{7db8ff} \\ 
        \hline
    \end{tabular}

    \label{tab:colorlegend}
\end{table}
\subsection{FaaS Results}
\begin{figure}[htbp]
  \centering
  \includegraphics[width=.95\linewidth]{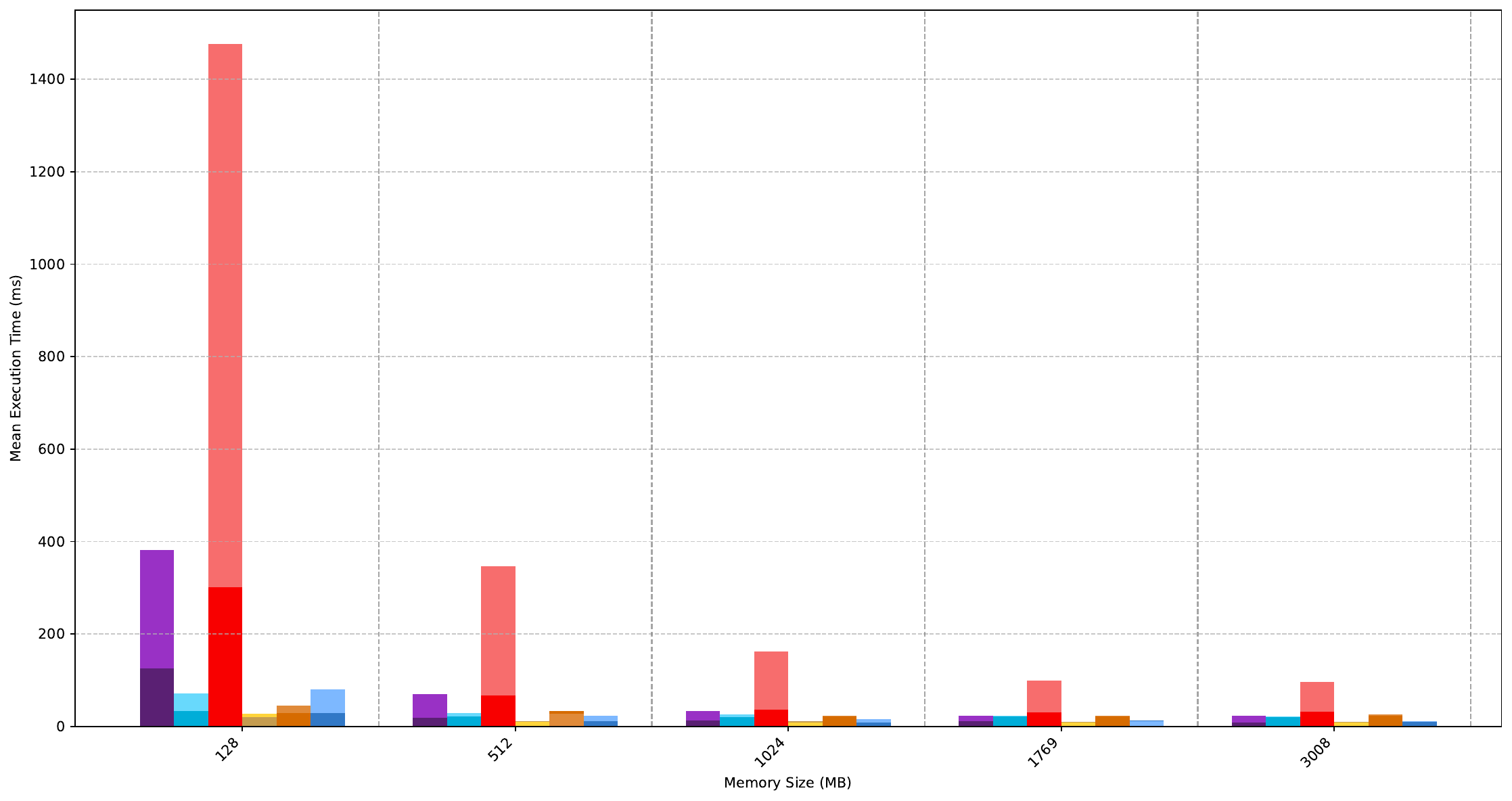}
  \vspace{1mm}
  \textbf{(a)}
  \includegraphics[width=.95\linewidth]{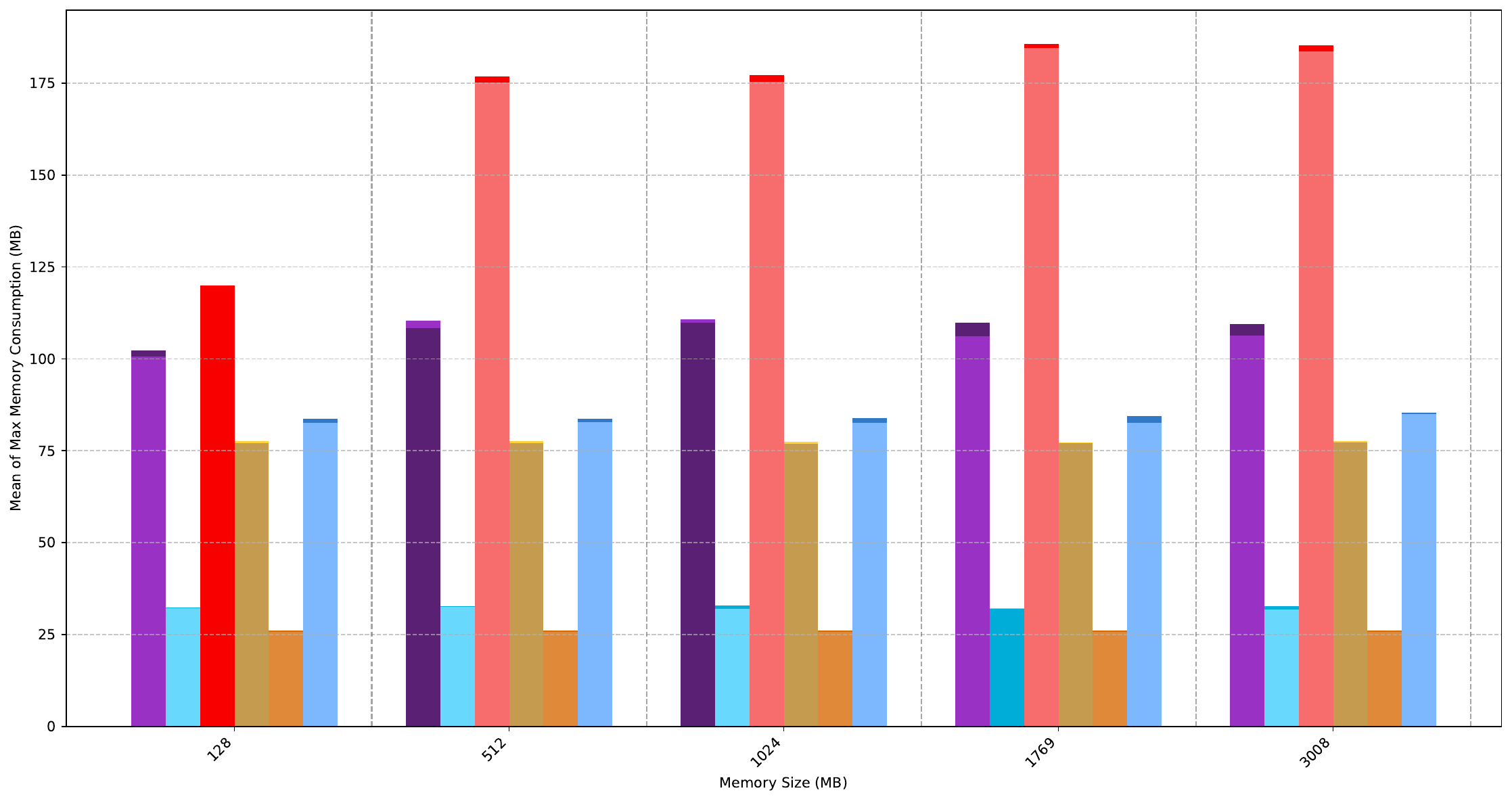}
  \vspace{1mm}
  \textbf{(b)}
  \includegraphics[width=.95\linewidth]{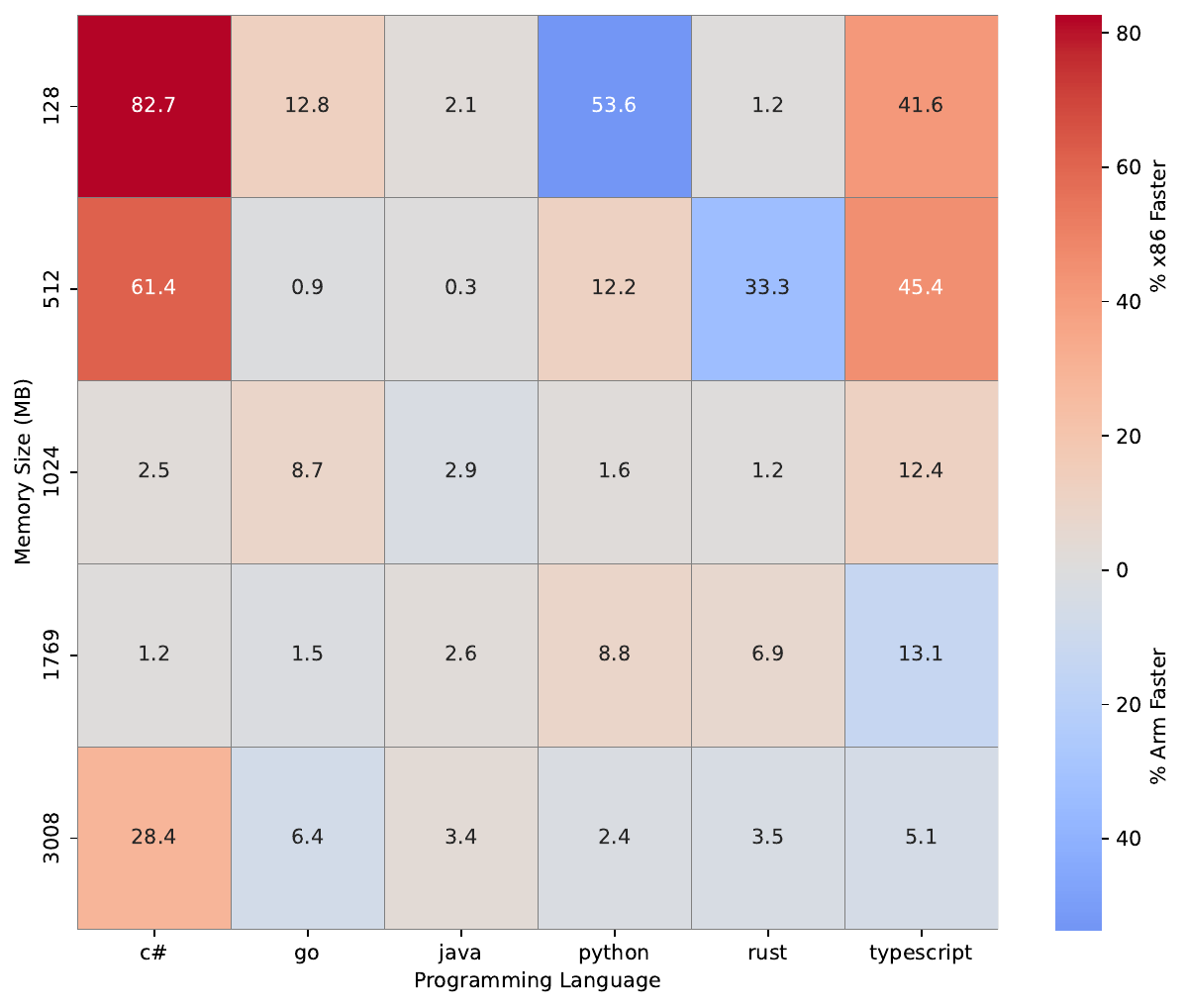}
  \vspace{1mm}
  \textbf{(c)}
  \caption{\textbf{FaaS SHA256 HMAC Generation Metrics. }
   (a) Mean Execution Time (ms) by memory size, (b) Mean of max memory (MB) consumption by memory size, (c) Heat Map of Warm execution speed percentage difference between x86\_64 and Arm64.}
  \label{fig:faas-sha256}
\end{figure}

\begin{figure*}[htbp]
  \centering
  \begin{minipage}[c]{0.75\linewidth}
    \centering
    \includegraphics[width=\linewidth]{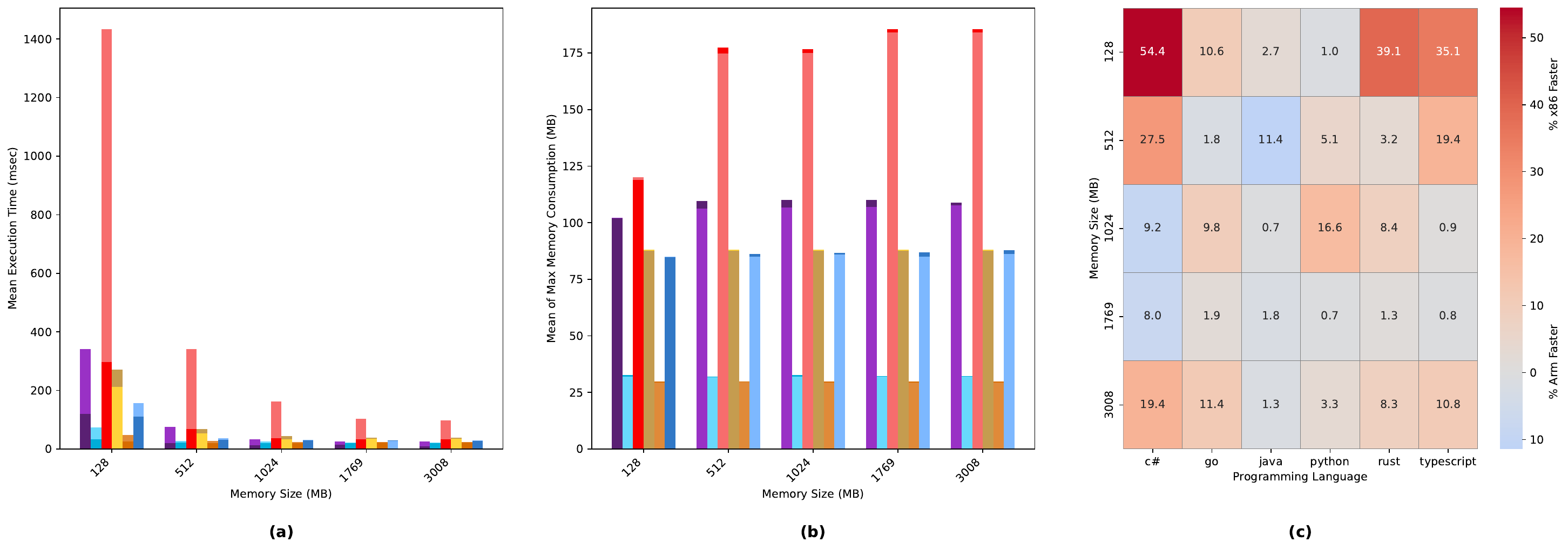}\\[0.5ex]
  \end{minipage}
  \hfill
  \\
  \vspace{1em} 

  \begin{minipage}[c]{0.75\linewidth}
    \centering
    \includegraphics[width=\linewidth]{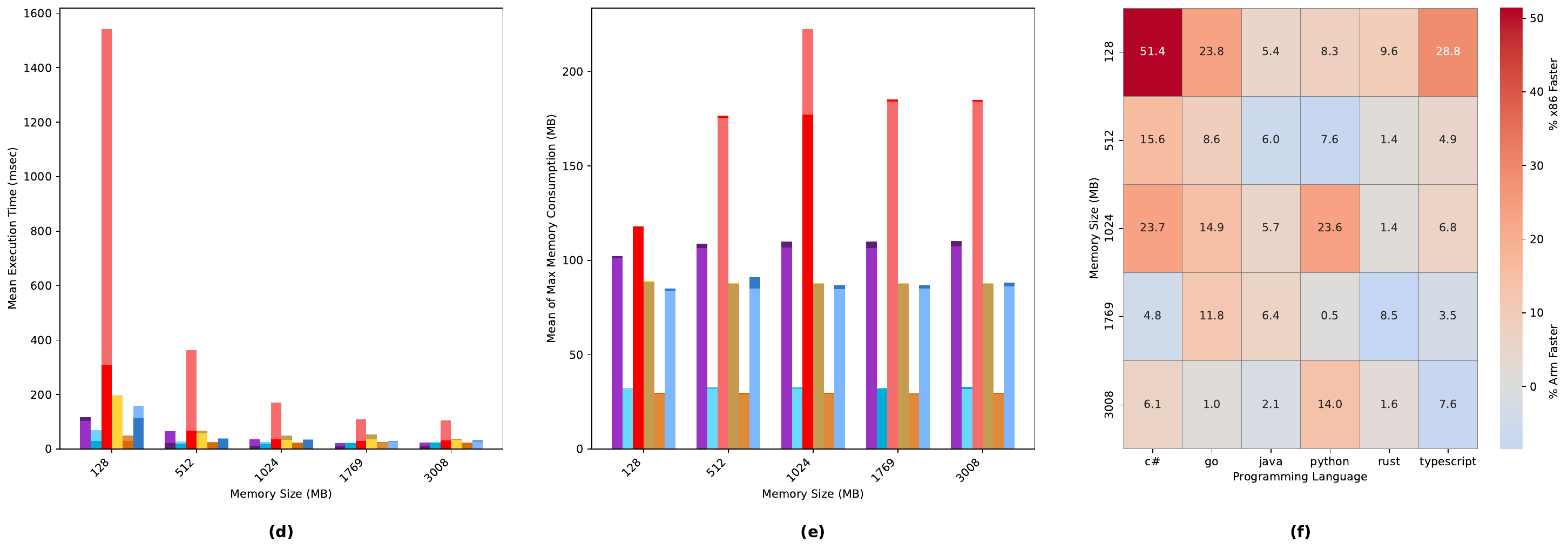}\\[0.5ex]
  \end{minipage}

  \caption{%
    \textbf{FaaS AES256 Encrypt/Decrypt Metrics.} 
    (a)--(c) show Encrypt metrics: (a) Mean execution time (ms) by memory size, 
    (b) Mean of max memory (MB) consumption by memory size, 
    (c) Heat map of warm execution speed percentage difference between x86\_64 and Arm64. 
    (d)--(f) show the corresponding Decrypt metrics.
  }
  \label{fig:faas-aes256-combined}
\end{figure*}

\begin{figure*}[htbp]
  \centering
  \begin{minipage}[c]{0.75\linewidth}
    \centering
    \includegraphics[width=\linewidth]{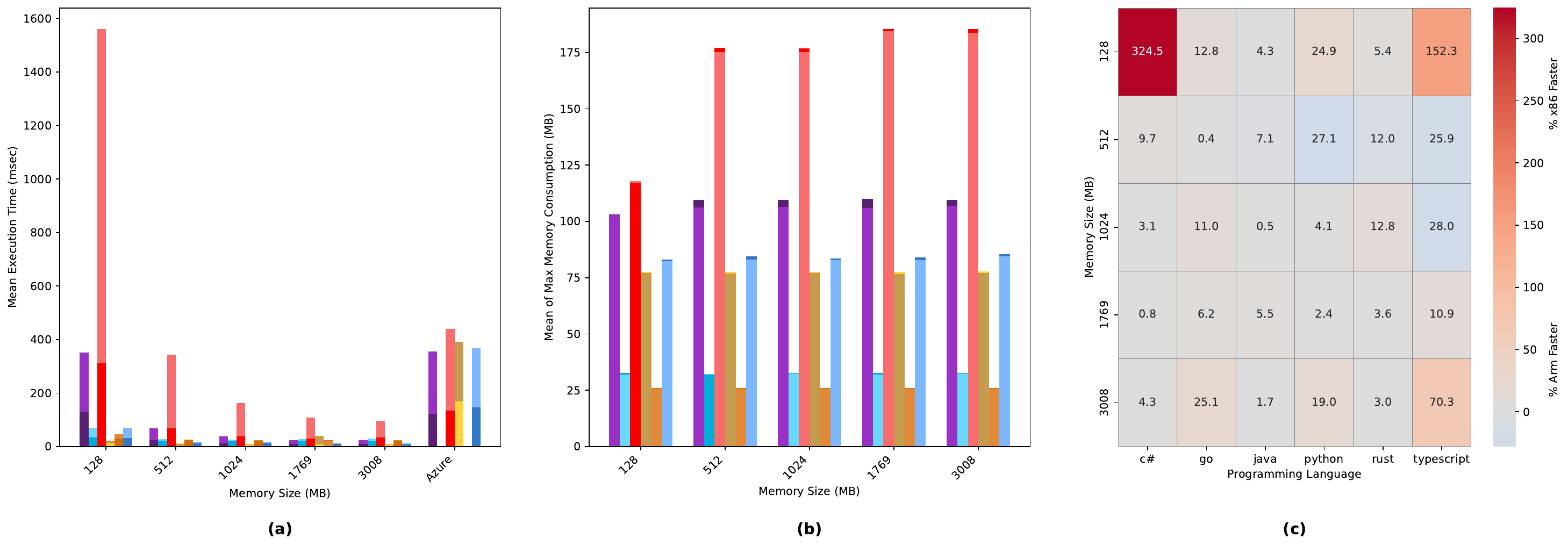}\\[0.5ex]
  \end{minipage}
  \hfill
  \\
  \vspace{1em} 

  \begin{minipage}[c]{0.75\linewidth}
    \centering
    \includegraphics[width=\linewidth]{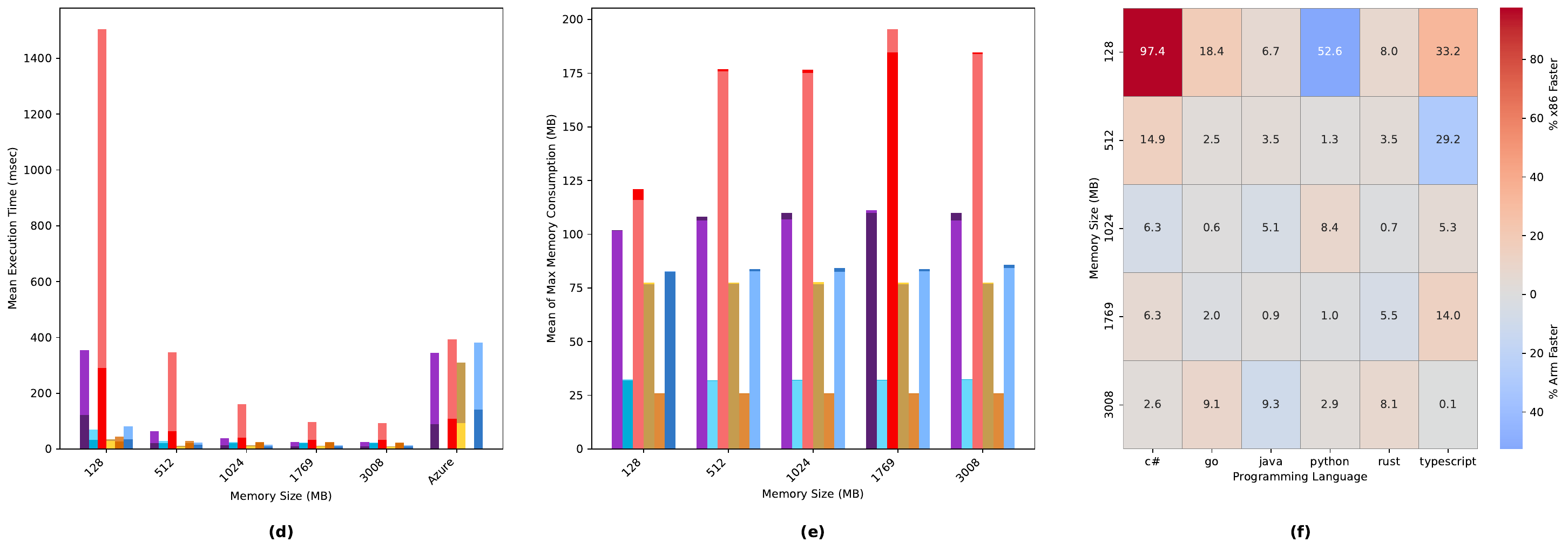}\\[0.5ex]
  \end{minipage}

  \caption{%
    \textbf{FaaS ECC256 Sign/Verify Metrics.}
    (a)--(c) show Sign metrics: (a) Mean execution time (ms) by memory size, 
    (b) Mean of max memory (MB) consumption by memory size, 
    (c) Heat map of warm execution speed percentage difference between x86\_64 and Arm64. 
    (d)--(f) show the corresponding Verify metrics.
  }
  \label{fig:faas-ecc256-combined}
\end{figure*}

\begin{figure*}[htbp]
  \centering
  \begin{minipage}[c]{0.75\linewidth}
    \centering
    \includegraphics[width=\linewidth]{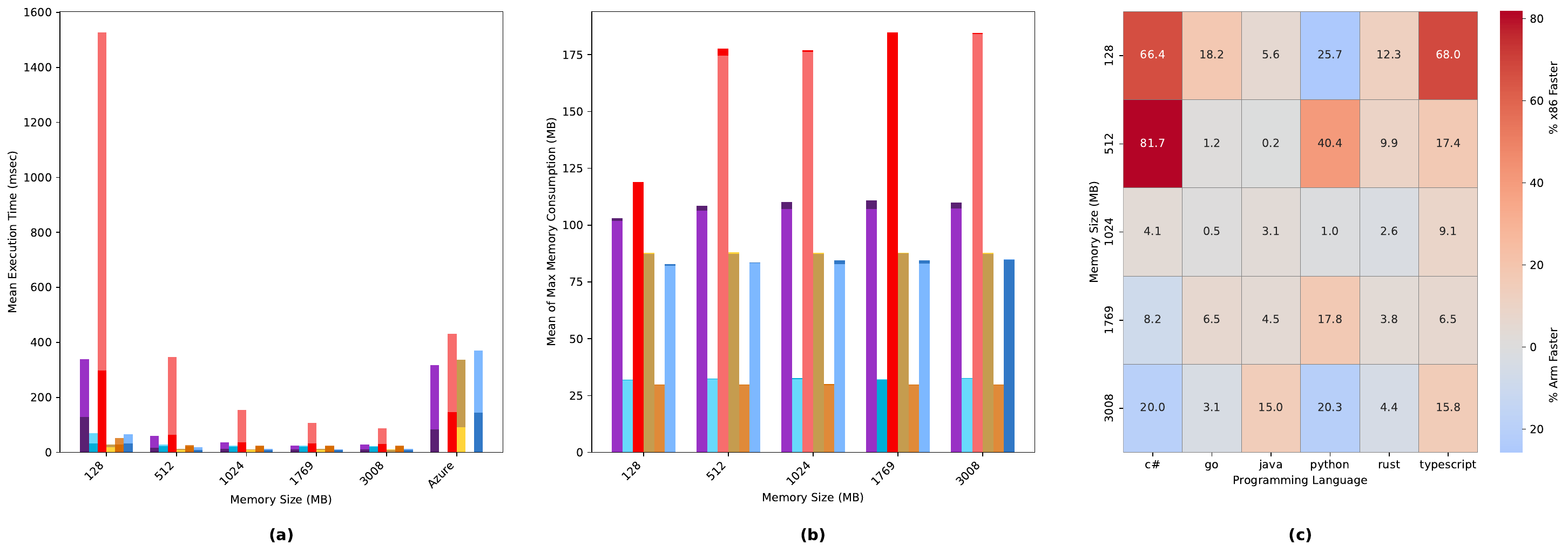}\\[0.5ex]
  \end{minipage}
  \hfill
  \\
  \vspace{1em} 

  \begin{minipage}[c]{0.75\linewidth}
    \centering
    \includegraphics[width=\linewidth]{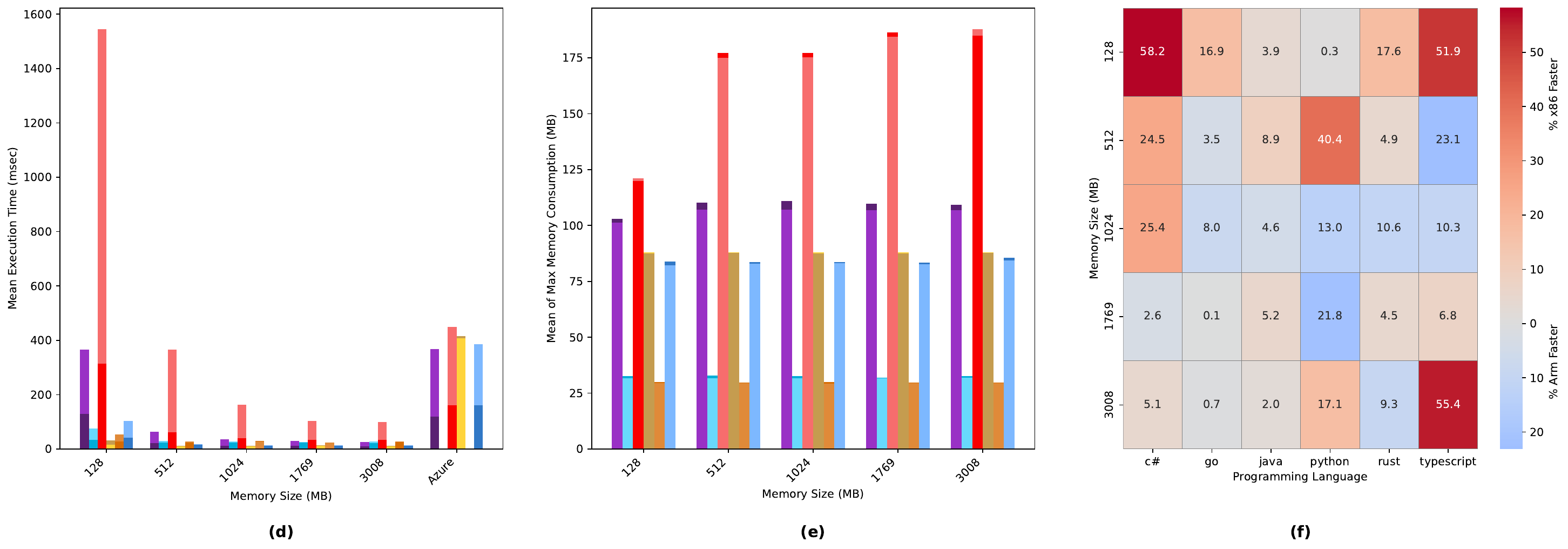}\\[0.5ex]
  \end{minipage}

  \caption{%
    \textbf{FaaS RSA2048 Encrypt/Decrypt Metrics.}
    (a)--(c) show the Encrypt metrics: (a) Mean execution time (ms) by memory size, 
    (b) Mean of max memory (MB) consumption by memory size, 
    (c) Heat map of warm execution speed percentage difference between x86\_64 and Arm64.
    (d)--(f) show the corresponding Decrypt metrics.
  }
  \label{fig:faas-rsa2048-combined}
\end{figure*}

This section presents our analysis of cryptographic workload performance across the AWS and Azure FaaS environments. We evaluated execution times, memory consumption, cost efficiency, and the influence of programming language selection on performance. For a comprehensive breakdown of cryptographic workload metrics based on language, memory size, architecture, and start type, refer to Figures \ref{fig:faas-sha256} - \ref{fig:faas-rsa2048-combined}. Higher bit-size metrics is provided in the github repository. The stacked bar charts illustrate the variations in time and memory consumption between cold and warm starts. For the workloads, bit size does not significantly impact execution time but instead affects memory consumption, increasing it by approximately 5–10\% relative to the workload's bit size.

\begin{table}[htbp]
\centering
\caption{FaaS - Fastest Execution Configurations of Cryptographic Workloads}
\label{tab:faas-fastest-workloads}
\resizebox{\columnwidth}{!}{%
    \begin{tabular}{l l c c c c c c}
        \toprule
        \textbf{Workload} & \textbf{Start Type} & 
        \makecell{\textbf{Mean Execution} \\ \textbf{Time (ms)}} & 
        \makecell{\textbf{Max Mean Memory} \\ \textbf{Consumption (MB)}} &
        \textbf{Architecture} & 
        \makecell{\textbf{Programming} \\ \textbf{Language}} & 
        \makecell{\textbf{Memory Size} \\ \textbf{(MB)}} & 
        \makecell{\textbf{Mean Cost} \\ \textbf{(USD)}} \\
        \midrule
        \multirow{2}{*}{AES256 Decrypt} 
         & Cold & 22.81 & 106.50 & x86\_64 & C\# & 1769 & 6.57e-07 \\
         & Warm & 9.92 & 121.47 & ARM64 & C\# & 1769 & 2.28e-07 \\
        \specialrule{0.2pt}{0pt}{0pt}
        \multirow{2}{*}{AES256 Encrypt} 
         & Cold & 22.53 & 32.10 & x86\_64 & Go & 3008 & 1.10e-06 \\
         & Warm & 9.62 & 108.77 & x86\_64 & C\# & 3008 & 4.71e-07 \\
        \specialrule{0.2pt}{0pt}{0pt}
        \multirow{2}{*}{ECC256 Sign} 
         & Cold & 9.76 & 79.80 & ARM64 & Python & 3008 & 3.83e-07 \\
         & Warm & 7.97 & 77.67 & x86\_64 & Python & 1769 & 2.29e-07 \\
        \specialrule{0.2pt}{0pt}{0pt}
        \multirow{2}{*}{ECC256 Verify} 
         & Cold & 9.34 & 79.43 & ARM64 & Python & 1024 & 1.24e-07 \\
         & Warm & 8.46 & 77.43 & x86\_64 & Python & 512 & 7.03e-08 \\
        \specialrule{0.2pt}{0pt}{0pt}
        \multirow{2}{*}{ECC384 Sign} 
         & Cold & 10.45 & 76.97 & x86\_64 & Python & 512 & 8.68e-08 \\
         & Warm & 8.40 & 77.57 & x86\_64 & Python & 512 & 6.97e-08 \\
        \specialrule{0.2pt}{0pt}{0pt}
        \multirow{2}{*}{ECC384 Verify} 
         & Cold & 9.52 & 79.57 & ARM64 & Python & 1024 & 1.27e-07 \\
         & Warm & 8.89 & 83.40 & x86\_64 & TypeScript & 1769 & 2.56e-07 \\
        \specialrule{0.2pt}{0pt}{0pt}
        \multirow{2}{*}{RSA2048 Decrypt} 
         & Cold & 11.10 & 87.67 & x86\_64 & Python & 3008 & 5.44e-07 \\
         & Warm & 9.61 & 89.93 & ARM64 & Python & 1769 & 2.21e-07 \\
        \specialrule{0.2pt}{0pt}{0pt}
        \multirow{2}{*}{RSA2048 Encrypt} 
         & Cold & 9.15 & 87.43 & x86\_64 & Python & 3008 & 4.48e-07 \\
         & Warm & 8.30 & 87.93 & x86\_64 & Python & 1769 & 2.39e-07 \\
        \specialrule{0.2pt}{0pt}{0pt}
        \multirow{2}{*}{RSA3072 Decrypt} 
         & Cold & 13.27 & 89.03 & ARM64 & Python & 3008 & 5.20e-07 \\
         & Warm & 11.00 & 89.67 & ARM64 & Python & 3008 & 4.31e-07 \\
        \specialrule{0.2pt}{0pt}{0pt}
        \multirow{2}{*}{RSA3072 Encrypt} 
         & Cold & 9.78 & 87.50 & x86\_64 & Python & 1024 & 1.63e-07 \\
         & Warm & 8.19 & 90.63 & ARM64 & Python & 512 & 5.49e-08 \\
        \specialrule{0.2pt}{0pt}{0pt}
        \multirow{2}{*}{RSA4096 Decrypt} 
         & Cold & 16.21 & 87.73 & x86\_64 & Python & 1024 & 2.71e-07 \\
         & Warm & 14.54 & 88.00 & x86\_64 & Python & 512 & 1.21e-07 \\
        \specialrule{0.2pt}{0pt}{0pt}
        \multirow{2}{*}{RSA4096 Encrypt} 
         & Cold & 9.42 & 87.53 & x86\_64 & Python & 3008 & 4.61e-07 \\
         & Warm & 8.03 & 88.00 & x86\_64 & Python & 1769 & 2.31e-07 \\
        \specialrule{0.2pt}{0pt}{0pt}
        \multirow{2}{*}{SHA256} 
         & Cold & 9.17 & 78.97 & ARM64 & Python & 1024 & 1.22e-07 \\
         & Warm & 8.09 & 77.33 & x86\_64 & Python & 1769 & 2.33e-07 \\
        \specialrule{0.2pt}{0pt}{0pt}
        \multirow{2}{*}{SHA384} 
         & Cold & 8.67 & 77.20 & x86\_64 & Python & 3008 & 4.25e-07 \\
         & Warm & 7.74 & 77.47 & x86\_64 & Python & 1769 & 2.23e-07 \\
        \bottomrule
    \end{tabular}
}
\end{table}

\begin{table}[htbp]
\centering
\caption{FaaS - Most Cost Efficient Configurations Cryptographic Workloads}
\label{tab:faas-cost-workloads}
\resizebox{\columnwidth}{!}{%
    \begin{tabular}{l l c c c c c c}
        \toprule
        \textbf{Workload} & \textbf{Start Type} & 
        \makecell{\textbf{Mean Execution} \\ \textbf{Time (ms)}} & 
        \makecell{\textbf{Max Mean Memory} \\ \textbf{Consumption (MB)}} &
        \textbf{Architecture} & 
        \makecell{\textbf{Programming} \\ \textbf{Language}} & 
        \makecell{\textbf{Memory Size} \\ \textbf{(MB)}} & 
        \makecell{\textbf{Mean Cost} \\ \textbf{(USD)}} \\
        \midrule
        \multirow{2}{*}{AES256 Decrypt}   
         & Cold & 48.50 & 26.90 & ARM64   & Rust   & 128 & 8.25e-08 \\
         & Warm & 32.10 & 27.30 & ARM64   & Rust   & 128 & 5.45e-08 \\
        \specialrule{0.2pt}{0pt}{0pt}
        \multirow{2}{*}{AES256 Encrypt} 
         & Cold & 45.50 & 26.80 & ARM64   & Rust   & 128 & 7.74e-08 \\
         & Warm & 25.15 & 29.80 & X86\_64 & Rust  & 128 & 5.28e-08 \\
        \specialrule{0.2pt}{0pt}{0pt}
        \multirow{2}{*}{ECC256 Sign} 
         & Cold & 11.40 & 78.77 & ARM64   & Python & 512 & 7.66e-08 \\
         & Warm & 12.57 & 77.47 & X86\_64 & Python & 128 & 2.64e-08 \\
        \specialrule{0.2pt}{0pt}{0pt}
        \multirow{2}{*}{ECC256 Verify} 
         & Cold & 22.00 & 76.80 & ARM64   & Python & 128 & 3.73e-08 \\
         & Warm & 13.49 & 77.00 & ARM64   & Python & 128 & 2.29e-08 \\
        \specialrule{0.2pt}{0pt}{0pt}
        \multirow{2}{*}{ECC384 Sign} 
         & Cold & 10.94 & 80.27 & ARM64   & Python & 512 & 7.33e-08 \\
         & Warm & 13.00 & 77.70 & X86\_64 & Python & 128 & 2.72e-08 \\
        \specialrule{0.2pt}{0pt}{0pt}
        \multirow{2}{*}{ECC384 Verify} 
         & Cold & 10.53 & 79.07 & ARM64   & Python & 512 & 7.05e-08 \\
         & Warm & 12.79 & 77.50 & X86\_64 & Python & 128 & 2.69e-08 \\
        \specialrule{0.2pt}{0pt}{0pt}
        \multirow{2}{*}{RSA2048 Decrypt} 
         & Cold & 12.80 & 87.73 & X86\_64 & Python & 512 & 1.06e-07 \\
         & Warm & 15.26 & 86.37 & ARM64   & Python & 128 & 2.59e-08 \\
        \specialrule{0.2pt}{0pt}{0pt}
        \multirow{2}{*}{RSA2048 Encrypt} 
         & Cold & 11.40 & 89.90 & ARM64   & Python & 512 & 7.66e-08 \\
         & Warm & 13.33 & 86.23 & ARM64   & Python & 128 & 2.26e-08 \\
        \specialrule{0.2pt}{0pt}{0pt}
        \multirow{2}{*}{RSA3072 Decrypt} 
         & Cold & 32.07 & 85.90 & ARM64   & Python & 128 & 5.43e-08 \\
         & Warm & 17.83 & 87.97 & X86\_64 & Python & 128 & 3.75e-08 \\
        \specialrule{0.2pt}{0pt}{0pt}
        \multirow{2}{*}{RSA3072 Encrypt} 
         & Cold & 11.72 & 88.73 & ARM64   & Python & 512 & 7.85e-08 \\
         & Warm & 8.19  & 90.63 & ARM64   & Python & 512 & 5.49e-08 \\
        \specialrule{0.2pt}{0pt}{0pt}
        \multirow{2}{*}{RSA4096 Decrypt} 
         & Cold & 32.07 & 85.80 & ARM64   & Python & 128 & 5.45e-08 \\
         & Warm & 17.60 & 87.97 & X86\_64 & Python & 128 & 3.70e-08 \\
        \specialrule{0.2pt}{0pt}{0pt}
        \multirow{2}{*}{RSA4096 Encrypt} 
         & Cold & 11.63 & 88.87 & ARM64   & Python & 512 & 7.79e-08 \\
         & Warm & 13.17 & 86.40 & ARM64   & Python & 128 & 2.24e-08 \\
        \specialrule{0.2pt}{0pt}{0pt}
        \multirow{2}{*}{SHA256} 
         & Cold & 11.31 & 79.50 & ARM64   & Python & 512 & 7.58e-08 \\
         & Warm & 12.50 & 77.00 & ARM64   & Python & 128 & 2.13e-08 \\
        \specialrule{0.2pt}{0pt}{0pt}
        \multirow{2}{*}{SHA384} 
         & Cold & 10.30 & 77.13 & X86\_64 & Python & 512 & 8.55e-08 \\
         & Warm & 12.88 & 77.67 & X86\_64 & Python & 128 & 2.70e-08 \\
        \bottomrule
    \end{tabular}%
}
\end{table}

Interestingly, for a majority of cryptographic operations on AWS Lambda, Python 3.11 consistently delivered the fastest execution times and often proved to be the most cost-efficient language. Rust was a close competitor, typically lagging Python by less than 5\% in raw performance; however, Rust excelled in memory usage by consuming significantly less peak memory, often multiple times lower than Python, making it especially attractive for workloads where memory footprint is a concern. This performance difference could be attributed to the size of the file input (4 KB).

In contrast, on Azure Functions, C\# exhibited superior performance in most cryptographic workloads, likely owing to the tight integration of the .NET runtime within the Azure platform. This gives C\#-based functions a notable edge in execution speed over other runtimes, particularly when sufficient memory resources are allocated to avoid throttling and garbage-collection delays. As a result, developers prioritizing minimal latency on Azure could capitalize on the strong synergy between the .NET runtime and Azure’s dynamic memory environment, whereas those seeking the lowest cost footprint or needing cross-platform consistency and memory specification might still find AWS Python, Go, or Java workload implementations advantageous.

\subsubsection{Architecture} 
As seen in Table \ref{tab:faas-cost-workloads}, x86\_64 functions consistently delivered superior warm-start performance, often shaving additional milliseconds off the execution time once the runtime was already initialized. This trend was particularly pronounced for higher-memory configurations (e.g., 512 MB or more), where x86\_64 could fully leverage the extra resources to handle cryptographic operations without significant overhead. Conversely, ARM64 stood out in cold-start scenarios, showing notably lower cold-start latencies across virtually all tested workloads. For instance, RSA3072 Decrypt and RSA4096 Decrypt on ARM64 at 128 MB memory settings both demonstrated faster cold-start times (reflected in lower mean execution times in the table), suggesting that ARM64’s reduced initialization overhead can be advantageous for workloads where frequent scaling or idle periods make cold-start delays more critical.

\subsubsection{Memory Configuration}
Lower memory configurations, particularly at 128 MB, consistently resulted in the most economical cost profile (refer to Table \ref{tab:faas-cost-workloads}), with invocation costs frequently falling into the order of $10^{-8}$ USD per execution. However, this economic advantage is often counterbalanced by increased execution latency, which is especially pronounced in cryptographic operations with complex or larger key sizes (e.g., RSA2048 and RSA3072). Conversely, higher memory allocations (512 MB or greater) typically delivered a more balanced tradeoff between execution time and cost, notably under steady-state (warm-start) scenarios. This indicates that, while minimal memory allocation achieves cost efficiency, organizations should carefully evaluate whether the associated latency penalties align with their workload requirements.

Moreover, Azure's dynamic memory allocation strategy, although convenient for handling workloads with unpredictable demands, may introduce unintended performance degradation. Automatic resource adjustments can result in frequent memory allocation shifts, leading to performance fluctuations, particularly in consistently demanding or latency-sensitive cryptographic operations. In contrast, AWS Lambda’s explicit memory configuration grants organizations precise control, enabling tailored optimization of performance and cost efficiency. Hence, organizations prioritizing consistent, predictable performance and cost efficiency, especially critical in secure cryptographic contexts may benefit more from AWS Lambda's granular configuration capabilities, despite the potential increase in management complexity.

\subsubsection{Execution Context}
Cold starts incur execution-time penalties, underscoring the importance of workload scheduling to maintain warm instances. A key area of divergence between AWS and Azure centered on cold-start versus warm-start behavior. As shown in Table \ref{tab:faas-fastest-workloads}, AWS x86\_64 functions generally achieved better warm-start performance, whereas ARM64 often excelled in cold-start efficiency, suggesting that ARM64’s lighter initialization overhead led to reduced startup latency. In Azure Functions, cold-start impacts were also evident—particularly with languages like C\#. On both platforms, optimizing cold-start performance proved critical for time-sensitive cryptographic tasks, underscoring the importance of maintaining “warm” instances in higher-throughput scenarios. AWS Lambda exhibited a lower variance in execution times compared to Azure Functions, which showed significant fluctuations, particularly during cold-start executions. Ultimately, runtime selection remains highly dependent on the specific requirements of each cryptographic use case, including memory constraints, cost constraints, and execution-time requirements.

\subsubsection{Programming Language}
Tables \ref{tab:faas-fastest-workloads} and \ref{tab:faas-cost-workloads} illustrate how the choice of language runtime can significantly affect both performance and resource consumption. On AWS, Python frequently emerged as the top performer for more demanding cryptographic workloads, whereas Azure favored C\#, leveraging .NET runtime integration for lower execution times. Rust and Go, although slightly behind Python in raw execution times, demonstrated distinct advantages in their significantly reduced memory footprints—often using less than half the memory required by Python or C\#. Specifically, Rust was particularly noteworthy for workloads such as AES256 encryption and decryption, combining near-best execution speeds with exceptionally low memory usage. Such efficiency positions Rust as an excellent candidate for cost-sensitive deployments or resource-constrained environments.

TypeScript (via Node.js) often rivaled Python’s execution times for cryptographic tasks, albeit at a higher resource usage. Java, in contrast, consistently exhibited the weakest performance in terms of both execution time and memory consumption across cryptographic tasks in these tests. The inherent overhead associated with JVM initialization, particularly pronounced during cold starts, significantly impacted its efficiency. Consequently, Java emerges as less favorable for performance-critical cryptographic functions in FaaS environments unless existing systems demand JVM compatibility. Ultimately, developers must weigh factors such as platform-specific optimizations, team familiarity, and cost-performance balance.

\section{Conclusion and Future work}
\label{Conclusion}

Overall, this study used microbenchmarking to conduct a multi-language, cross-architecture evaluation of cryptographic workload performance across FaaS on AWS and Azure. We studied how to conduct microbenchmarking of cryptographic workloads on AWS Lambda and Azure Functions via their respective KMS services. The key determinants of cryptographic workload performance (RQ1) in an FaaS context: low language runtime overhead, efficient resource usage, and increased allocation of memory and vCPU, which decrease execution time. Azure is constrained in its capabilities, however, as it does not support some runtimes that AWS can. For RQ2, tuning instance types (e.g., leveraging ARM-based t4g.medium or x86 t2.xlarge), right-sizing vCPU and memory allocations, and selecting a lightweight compiled runtime (such as Rust) emerged as the most impactful optimizations. Ultimately, each environment’s performance/cost trade-offs highlight the importance of holistic testing and benchmarking to identify the optimal configuration for specific cryptographic workloads across different cloud providers. RPM systems can leverage FaaS cryptographic benchmarking insights, through optimized runtimes, memory/vCPU tuning, and secure KMS integration to mitigate latency, key management, and data interception threats while ensuring real-time patient monitoring. Looking ahead, future research can extend this benchmarking framework by exploring additional workload configurations and cloud environments. One promising avenue is testing higher AWS Lambda memory allocations, as AWS supports configurations up to 10,240 MB, which may impact performance for memory-intensive cryptographic operations.

\bibliographystyle{IEEEtran}
\bibliography{References}

\end{document}